\documentclass[pra,twocolumn,amsmath,amssymb]{revtex4}

\usepackage{url}
\usepackage{amsmath,amssymb}
\usepackage{mathrsfs}
\usepackage{ifthen}

\usepackage{ifpdf}
\ifpdf
\usepackage[pdftex]{hyperref}
\fi

\newcommand\ket[1]{{ |{#1} \rangle }}

\newcommand{\ignore}[1]{}

\newcommand{\qedsymb}{\hfill{\rule{2mm}{2mm}}}

\begin{document}

\title{Correctable noise of quantum error correcting codes under adaptive concatenation}
\author{Jesse Fern}

\affiliation{Berkeley Quantum Information Center, Department of Mathematics, University of California, Berkeley, California, 94720
}

\email{jesse@math.berkeley.edu}

\begin{abstract}
We examine the transformation of noise under a quantum error correcting code (QECC) concatenated repeatedly with itself, by analyzing the effects of a quantum channel after each level of concatenation using recovery operators that are optimally adapted to use error syndrome information from the previous levels of the code. We use the Shannon entropy of these channels to estimate the thresholds of correctable noise for QECCs and find considerable improvements under this adaptive concatenation. Similar methods could be used to increase quantum fault tolerant thresholds.
\end{abstract}
\maketitle

Decoherence is a significant problem for implementation of quantum computing. Quantum error correction \cite{Shor1,KL} mitigates decoherence and gate errors by introducing redundancy. In particular, for many codes, one qubit is encoded into $n$ qubits, and each of those encoded into $n$ qubits, so that after $j$ levels of concatenation, $n^j$ qubits are used.  In previous work, concatenated quantum codes were analyzed using a channel map formalism to evaluate their performance and to determine the threshold error probability for perfect fidelity in the infinite-concatenation limit\cite{Rahn:02a,FKSS}.
In order to optimize the recovery operators at each level of the code, we develop here an adaptive concatenation approach, which yields significantly larger values for the thresholds of correctable noise. This work extends previous work \cite{Poulin}, by formulating it in the channel superoperator representation, which allows for arbitrary quantum channels. We introduce the idea of using the Shannon entropy of ensembles of channels to calculate the thresholds of correctable noise; we use this to find thresholds for two quantum error correcting codes under two different types of noise.

\section{Basic notation}
The $n$ qubit Pauli operators are $\mathcal{P}_n = \{I,X,Y,Z\}^{\otimes n}$. We use the notation that $XY = X \otimes Y$. Each pair of Pauli operators $\sigma, \sigma'$ either commutes or anticommutes,
where $\sigma \in \mathcal{P}_n$. We let the quantity $\eta(\sigma, \sigma')$ be $1$ if they commute, $-1$ if they anticommute.

We write a density matrix on $n$ qubits as $\rho = \frac{1}{2^n} \sum_{\sigma \in \mathcal{P}_n} c_\sigma \sigma$.
A channel is a map from density matrices to density matrices, and can be written as $\$(\rho) = \sum_i A_i \rho A_i^{\dagger}$, where the $A_i$ are Kraus operators that act on the state. This channel has the matrix representation $\mathcal{N} = \sum_i \mathcal{O}(A_i)$, where $\mathcal{O}(M)=M \otimes M^{\dagger T}$ and $\mathcal{N}$ is a matrix that acts on the vector form of $\rho$ in the standard basis.
By looking at how the $4^n$ Pauli operators $\sigma \in \mathcal{P}_n$ perform under the map $\$(\sigma)$, the channel can be written as a $4^n$ by $4^n$ superoperator in the Pauli basis. Because channels are trace-preserving and hermitian-preserving, this implies that one-qubit channels written in the Pauli basis have real entries and the first row is $\mathcal{N}_{I \sigma} = \delta_{I, \sigma}$. On one-qubit, this is
\begin{equation*}
\label{noise}
\mathcal{N}^{(1)}=
\begin{bmatrix}
1 & 0 & 0 & 0 \cr
N_{XI} & N_{XX} & N_{XY} & N_{XZ} \cr
N_{YI} & N_{YX} & N_{YY} & N_{YZ} \cr
N_{ZI} & N_{ZX} & N_{ZY} & N_{ZZ}
\end{bmatrix}.
\end{equation*}

A quasi-density matrix $\rho$ is an unnormalized density matrix. It can be normalized to a density matrix $\frac{1}{p_s} \rho$, where $p_s = \text{tr } \rho = c_I$. Just as a density matrix can be written as a sum of quasi-density matrices, a channel can be written as a sum of quasi-channels \cite{FKSS}. Quasi-channels map density matrices to quasi-density matrices, and therefore do not have to preserve trace. If $\mathcal{N}$ is a quasi-channel and $\rho$ is a density matrix, then the probability of the quasi-channel occurring is $p_c = \text{tr} (\mathcal{N} \rho) = \sum_{\sigma} N_{I \sigma} c_\sigma$, which depends on the state $\rho$ unless $\mathcal{N}_{I \sigma} = p_c \delta_{I, \sigma}$, in which case it gives the channel $\frac{1}{p_c} \mathcal{N} = \frac{1}{\mathcal{N}_{II}} \mathcal{N}$ with probability $p_c$.

\section{Overview of Stabilizer codes}

An $[[n,k,d]]$ quantum stabilizer code encodes $k$ logical qubits into $n$ physical qubits, and has a distance $d$. Typically, a quantum stabilizer code is given in terms of $n-k$ generators $g_i$, and $4^k$ encoded Pauli operators $\overline{\sigma}$.

The $g_i$ generate the stabilizer group $S$, which is isomorphic to $Z_2^{n-k}$. The $k$ logical qubits encode into the $2^k$-dimensional code space $C_S$, which has the property that if $\ket{\psi} \in C_S$ and $s \in S$, then $s \ket{\psi} = \ket{\psi}$. The elements of the set $\mathcal{C}(S) \subset \mathcal{P}_n$ are the $2^{n+k}$ Pauli operators which commute with $S$. They send states in $C_S$ to states in $C_S$. Each of the $4^k$ different equivalence classes of $\mathcal{C}(S) /S$ corresponds to one of the $4^k$ encoded logical Pauli operators $\overline{\sigma}$ that act on the logical qubits in the code space.

To perform error detection, we measure each of the $n-k$  generators $g_i$, projecting into either the $+1$ or $-1$ eigenspace for each generator $g_i$. This gives us a syndrome bit $\beta_i$, which is $0$ if the state is in the $+1$ eigenspace, and $1$ if it is in the $-1$ eigenspace. These $\beta_i$ form the $2^{n-k}$ possible error syndromes $\beta$. $\beta_0=0=(0,0,\ldots,0)$ is the syndrome corresponding to no error.

To recover from an error characterized by syndrome $\beta$, a recovery operator $R(\beta)$ must be chosen that returns the qubit to the code space. Since $g_i R(\beta) \ket{\psi} = R(\beta) \ket{\psi}$, then  $\eta(R(\beta), g_i) = \beta_i$ for all $i$.  The $2^{n+k}$ possible choices for $R(\beta)$ are all in the same equivalence class of $\mathcal{P}_n / \mathcal{C}(S)$.

\section{Channel map formalism}
We define a $4^n$ by $4^k$ encoding operator $\mathcal{E}$ that maps from a density matrix on $k$ qubits to one on $n$ qubits as
\begin{equation*}
 \rho \rightarrow \frac{1}{2^{n-k}} \mathcal{E} \rho.
\end{equation*}
This projects us into the code space of the code. In the Pauli basis, the $4^k$ columns of $\frac{1}{2^{n-k}} \mathcal{E}$ act as the logical Pauli operators in the code space, and $0$ outside the code space. They are defined as
\begin{align}
\label{Esigmaform}
\mathcal{E}_I = \prod_i(I+g_i) \quad \mathrm{and} \quad \mathcal{E}_{\sigma} = \overline{\sigma} \mathcal{E}_I.
\end{align}
Now that the logical qubits are encoded into the code space, assume that some noise $\mathcal{N}$ acts on the full $2^n$-dimensional space of the $n$ physical qubits of the code. The $g_i$ operators are measured, giving an error syndrome $\beta$. Some unitary recovery operator $R(\beta)$ is then chosen. We can write the error-correction process as
\begin{equation*}
\mathcal{T}_{\text{ideal}} = \sum_{\beta} \mathcal{O}(R(\beta)) \circ \mathcal{P}_{\beta} = \mathcal{P}_0 \circ \sum_{\beta}  \mathcal{O}(R(\beta))
\end{equation*}
where $\mathcal{O}(R(\beta))$ is the error recovery superoperator, $\mathcal{P}_{\beta}$ is a projection superoperator into the syndrome $\beta$ space, and $\mathcal{P}_0 = \frac{1}{2^{n-k}} \mathcal{E} \circ \mathcal{E}^T$ (transpose of $\mathcal{E}$ followed by $\mathcal{E}$) is a projection into the code space. $\mathcal{E}^t$ decodes back into the original $k$ qubit space. The whole process can then be written as
\begin{equation}
\label{generalcodingmap}
\mathcal{G} = \frac{1}{2^{n-k}} \mathcal{E}^t \circ \sum_{\beta} \mathcal{O}(R(\beta)) \circ \mathcal{N} \circ \mathcal{E}.
\end{equation}
This represents the noise on the logical qubits $\mathcal{G}$ in terms of the noise on the physical qubits $\mathcal{N}$.  We can decompose $\mathcal{G}$ as a sum
of contributions from each error syndrome:
\begin{equation}
\label{Gsyndpart}
\mathcal{G}=\sum_{\beta} \mathcal{G}^{R(\beta)}, \quad \mathcal{G}^U=  \frac{1}{2^{n-k}} \mathcal{E}^t \circ \mathcal{O}(U) \circ \mathcal{N} \circ \mathcal{E}.
\end{equation}

\section{Same noise on each qubit}
Often it is assumed that each qubit experiences the same noise, so $\mathcal{N} = \left(\mathcal{N}^{(1)} \right)^{\otimes n}$.
Previous work \cite{Rahn:02a,FKSS} showed that the case of a code $C$ where $k=1$ produces a map
$
\Omega^C : \mathcal{N}^{(1)} \rightarrow \sum_\beta \frac{1}{2^{n-k}} \mathcal{E}^t \circ  \mathcal{O}(R(\beta)) \circ \left( \mathscr{N}^{(1)} \right)^{\otimes n} \circ \mathcal{E}
$
from the physical one-qubit noise on each qubit to the logical one-qubit noise on the encoded code space. This is useful for analyzing a code concatenated with itself an arbitrary number of times. A logical qubit on one level of the code is treated as a physical qubit at the next of the code, allowing us to concatenate a code with itself many times. In particular, the code will correct noise if $\lim_{k \to \infty} {\Omega^C}^{\circ k}(\mathcal{N}^{(1)}) = I$ \cite{FKSS}. However, this assumes that the recovery operators do not depend on syndrome information from the previous levels of error correction. Furthermore, it does not give as high thresholds as the optimal adaptive concatenation described below.

\section{Diagonal noise}
A Pauli matrix $\sigma$ has the diagonal superoperator $\mathcal{O}(\sigma)$, where $\mathcal{O}(\sigma)_{\sigma' \sigma'} = \eta(\sigma, \sigma')$.
Suppose the noise is diagonal in the Pauli basis. Then the $\mathcal{G}^{U}$ from Eq. \ref{Gsyndpart} become $\mathcal{G}_{\sigma' \sigma'}^{\sigma}  = \frac{1}{2^{n-k}} \mathcal{E}_{\sigma'}^t \circ  \mathcal{O}(\sigma) \circ \mathcal{N} \circ \mathcal{E}_{\sigma'}$. Since the nonzero entries of the column $\mathcal{E}_{\sigma'}$ are $\overline{\sigma'} s$ where $s \in S$, it follows that
\begin{equation}
\label{diagmap}
\mathcal{G}_{\sigma', \sigma' }^{\sigma} = \frac{1}{2^{n-k}} \sum_{s \in S} \eta(\sigma, \overline{\sigma'} s) \mathcal{N}_{\overline{\sigma'} s,\overline{\sigma'} s}.
\end{equation}

If a quasi-channel is diagonal, it can be written as a sum of Pauli superoperators $\mathcal{N} = \sum_{\sigma} p_{\sigma} \mathcal{O}(\sigma)$, which corresponds to having each Pauli error $\sigma$ with probability $p_\sigma$. For a one-qubit quasi-channel with probability $p = \sum_\sigma p_\sigma$, the diagonal parts are
\begin{align}
\label{eq:diagchan}
[p,x,y,z] = p_I [1,1,1,1] + p_X [1,1,-1,-1] \nonumber \\
+ p_Y [1,-1,1,-1], + p_Z [1,-1,-1,1].
\end{align}
This yields the Pauli probabilities:
\begin{align}
\label{eq:pauliprob}
p_I = \frac{p+x+y+z}{4}
&& p_X = \frac{p+x-y-z}{4}\\
p_Y = \frac{p-x+y-z}{4}
&& p_Z = \frac{p-x-y+z}{4}. \nonumber
\end{align}

Since the information entropy (Shannon) in terms of Pauli errors with probabilities $q_i$ is $\sum_i h(q_i)$ where $h(x) = - x \log_2 x$, a quasi-channel gives a contribution of
%\begin{equation*}
$p \sum_\sigma h(\frac{p_\sigma}{p}) = -h(p) + \sum_\sigma h(p_\sigma)$
%\end{equation*}
to the entropy of a channel.

\section{Adaptive concatenation}
We now show how using syndrome information from the lowest levels of a code can be used to optimize the threshold of the code.

When we use the syndrome information, instead of having the logical encoded channel map written as a simple sum of channel contributions $\mathcal{G}=\sum_\beta G^{R(\beta)}$ as in Eq. \ref{Gsyndpart}, we represent this additional information as
$
\sum_\beta G^{R(\beta)} \otimes \beta
$.
If we perform error correction and ignore the syndrome information, we have simply $\mathcal{G}$. The probability of measuring a syndrome $\beta$
depends on the density matrix state $\rho$, and is
\begin{equation*}
p_{\beta} =\text{tr}(\mathcal{G}^{R(\beta)} \rho) = 2^n \sum_{\sigma} \mathcal{G}_{I \sigma}^{R(\beta)} c_{\sigma}.
\end{equation*}
If the first row of this quasi-channel in the Pauli basis is all zero except for the first term, i.e., $\mathcal{G}_{I \sigma}^{R(\beta)}= p\delta_{I \sigma}$, then $p_{\beta} = \mathcal{G}_{II}^{R(\beta)} = p$. If the syndrome $\beta$ is measured, the resulting noise is $\frac{1}{p_{\beta}} \mathcal{G}^{R (\beta)}$.

\section{Recovery optimization}
Given an $[[n,1,d]]$ quantum code, assume that the syndrome $\beta$ is measured. It has a corresponding equivalence class $\mathcal{P}_n / C(S)$. From this, we choose some representative element $r_\beta$  to return to the code space. This element differs from the optimal recovery operator $R(\beta)$ by a logical Pauli operator $\overline{\sigma}$.  The recovery operators for $\beta$ are therefore equivalent to the $4$ recovery operators $\overline{\sigma} r_\beta $, where $\overline{\sigma}$ is any one of the $4$ logical encoded Pauli operators. We then have the quasi-channel $\mathcal{G}^{r_\beta}$.

Now, for each of the $n$ blocks of the code, we pass this quasi-channel $\mathcal{N}_i$ onto the next level of the code, and have the resulting noise $\mathcal{N} = \bigotimes_i \mathcal{N}_i$. To optimize the recovery operator based upon this noise, either we can  find the optimal $\overline{\sigma} r_\beta$ at each level of the code, or we can apply a random recovery operator $r_\beta$ at each level of the code, and optimize at the end.
In the second case, since each block differed by a Pauli operator $\sigma_i$ from optimal, the resulting noise differs by $\otimes_i \sigma_i$ from optimal. This is equivalent to having the recovery operator $\otimes_i \sigma_i \circ r_\beta$ instead of $r_\beta$. This is $\overline{\sigma} \circ r_\alpha$ for some logical Pauli operator $\overline{\sigma}$ and some error syndrome $\alpha$, and therefore differs by a logical Pauli operator from the optimal recovery operator at the top level of the code.  %Optimization can be also delayed if the optimal $R(\beta)$ is not a Pauli operator, although this is more complicated.

While the first method may seem more intuitive, the second may be more computationally efficient, since determining the optimal recovery operator at any given level of the code is nontrivial. At the last level of the code, we will choose the encoded Pauli operator $\sigma$ that yields the channel closest to the identity channel, which corresponds to the maximal $p_\sigma$ from Eq. \ref{eq:pauliprob}.  If the resulting noise is very close to the identity channel, then we have established that very little logical noise is introduced at each step of a quantum computation, allowing long quantum computations to be performed. In practice, the noise at the end of the concatenation will often be close to diagonal even though the initial physical qubit noise was not.

Suppose we wish to calculate the exact optimized channel map of an $[[n,1,d]]$ quantum code after $j$ levels of concatenation with itself. There are $2^{n^j-1}$ syndromes.
Many of these syndromes may have the same channel components $\mathcal{G}^{R(\beta)}$, rendering it unnecessary to calculate all of them, reducing the complexity. However, for many codes it is still computationally too expensive to compute the optimized channel past the second level.
Nevertheless, the overhead involved with implementing the optimization is low because the encoded channel from one level is simply passed on to the qubits of the next level of the code. This leads to efficient Monte Carlo simulation.

\section{Example: The two qubit bit flip code}
It is useful to look first at the two qubit bit flip code because of its simplicity; it is the classical two bit repetition code. Its code space is spanned by the logical encoded states $\overline{\ket{0}}=\ket{00}$, and $\overline{\ket{1}} = \ket{11}$. The stabilizer group $S=\{II, ZZ\}$ is generated by the one generator $g_1 = ZZ$. Tab. \ref{tab:2qubit} lists the encoded Pauli operators and corresponding $\mathcal{E}_\sigma$ (Eq. \ref{Esigmaform}).
\begin{table}
\caption{\label{tab:2qubit} Encoded Pauli operators $\overline{\sigma}$ and encoding operator $\mathcal{E}_\sigma$ for the 2 qubit code.
}
\begin{ruledtabular}
\begin{tabular}{ccc}
$\sigma$ & $\overline{\sigma}$ & $\mathcal{E}_\sigma$ \cr
\hline
$I$ & $II$ & $II + ZZ$ \cr
$X$ & $XX$ & $XX - YY$ \cr
$Y$ & $XY$ & $XY + YX$ \cr
$Z$ & $IZ$ & $IZ + ZI$ \cr
\end{tabular}
\end{ruledtabular}
\end{table}
Using Eq. \ref{Gsyndpart}, we can calculate various channel map components $G^{R(\beta)}$. For example,
%\begin{align*}
$
\mathcal{G}^{IX}_{X,Z} = \frac{1}{2} (-N_{XX,IZ} + N_{XX,ZI} + N_{YY,IZ} - N_{YY,ZI})
%\end{align*}
$. In the case of diagonal noise, the channel map components  can be found directly from Eq. \ref{diagmap}.  These are given in Tab. \ref{tab:2qubitchannel}, using the shorthand $\mathcal{N}_{\sigma,\sigma} = \sigma$.
\begin{table}
\caption{\label{tab:2qubitchannel} $\mathcal{G}^{R(\beta)}$ for diagonal noise, using the shorthand $\sigma = N_{\sigma, \sigma}$.
}
\begin{ruledtabular}
\begin{tabular}{ccc}
$\beta$ & $R(\beta)$ & $\mathcal{G}^{R(\beta)}$ \cr
\hline
$0$ & $II$ & $\frac{1}{2}[II+ZZ, XX+YY, XY+YX, IZ+ZI]$ \cr
$1$ & $XI$ & $\frac{1}{2}[II-ZZ, XX-YY, XY-YX, IZ-ZI]$ \cr
$1$ & $IX$ & $\frac{1}{2}[II-ZZ, XX-YY, YX-XY, ZI-IZ]$
\end{tabular}
\end{ruledtabular}
\end{table}

Suppose now that we have the same diagonal noise $\mathcal{N}^{(1)} = [1,1,x,x]$ on each qubit, with $x > 0$.  We see from Eq. \ref{eq:diagchan} that this represents an $X$ error with probability $p_x = \frac{1-x}{2}$. From Eq. \ref{Gsyndpart},
\begin{align*}
\mathcal{G}^{II}& = [\frac{1+x^2}{2}, \frac{1+x^2}{2}, x, x] \\
\mathcal{G}^{IX} = \mathcal{G}^{XI}& = [\frac{1-x^2}{2}, \frac{1-x^2}{2}, 0, 0],
\end{align*}
and the total map for this noise is
\begin{equation*}
\Omega^{bf_2}[1,1,x,x]=
\mathcal{G} = \mathcal{G}^{II} + \mathcal{G}^{IX} = \mathcal{G}^{II} + \mathcal{G}^{XI} = [1,1,x,x].
\end{equation*}
It is evident that this code is not very useful at correcting bit flip errors, since it leaves the bit flip channel unchanged.

If the code is concatenated with itself, the resulting map on the bit flip channel is $\Omega^{bf_2} \circ \Omega^{bf_2} [1,1,x,x] = [1,1,x,x]$, which is again ineffective at correcting bit flip errors. The problem is that the recovery operators are also concatenated using this method. Let the recovery operators be $R=\{II, XI\}$. Then, for one of the syndromes, we apply the recovery operator $XXXI$ instead of the optimal $IIIX$. This syndrome has no error detected in the first block, an error detected in the second block, and an error detected at the top level of the code. The block noises are $\mathcal{N}_1 = \mathcal{G}^{II}$ and $\mathcal{N}_2 = \mathcal{G}^{XI}$. Applying the default $XI$ recovery operator at the top level of the code (total recovery operator of $XXXI$) gives a logical noise of
\begin{align*}
\mathcal{G}^{XI} = \frac{1}{2}[II-ZZ, XX-YY, XY-YX, IZ-ZI]\\
= \frac{1}{2}[\frac{1+x^2}{2} \frac{1-x^2}{2} - x 0, \frac{1+x^2}{2} \frac{1-x^2}{2} - x 0,\\
 \frac{1+x^2}{2} 0 - x \frac{1-x^2}{2}, \frac{1+x^2}{2} 0 - x \frac{1-x^2}{2}]\\
= \frac{1-x^4}{8}[1,1,-\frac{2x}{1+x^2},-\frac{2x}{1+x^2}].
\end{align*}
For this quasi-channel, $p_X > p_I$ and $p_Y=p_Z=0$ (see Eq. \ref{eq:pauliprob}), and so the optimal recovery operator differs by encoded $\overline{X}$ from this. Since $\overline{X}=XX$, this results in a recovery operator of $IX$ instead of $XI$ at the top level of the code (total recovery operator $IIIX$). This gives
\begin{equation*}
\mathcal{G}^{IX} = [1,1,-1,-1]\mathcal{G}^{XI} = \frac{1-x^4}{8}[1,1,\frac{2x}{1+x^2},\frac{2x}{1+x^2}].
\end{equation*}
The difference $\Delta = \mathcal{G}^{II} - \mathcal{G}^{XI}$ between these is
\begin{equation*}
\Delta =  [0,0,YX-XY,ZI-IZ] = [0,0,x\frac{1-x^2}{2},x\frac{1-x^2}{2}].
\end{equation*}
%\begin{align*}
%\Delta = \mathcal{G}^{IX} - \mathcal{G}^{XI} = [0,0,YX-XY,ZI-IZ] \\
%= [0,0,x\frac{1-x^2}{2},x\frac{1-x^2}{2}].
%\end{align*}
Since this was the only suboptimal case, we optimize the total $4$ qubit code by adding $\Delta$  to the unoptimized channel map $[1,1,x,x]$, yielding the optimized map
\begin{equation*}
\Omega^{bf_4}[1,1,x,x] = [1,1,\frac{3}{2}x - \frac{1}{2}x^3, \frac{3}{2}x-\frac{1}{2}x^3].
\end{equation*}
Since $x \rightarrow \frac{3}{2}x - \frac{1}{2}x^3$ converges to $1$ for $x > 0$, iterating this map now yields a useful code for correcting $X$ errors.

\begin{table}
\caption{\label{tab:depolarizing} Critical values for depolarizing channel $(p_X,p_Y,p_Z)=(p,p,p)$.}
\begin{ruledtabular}
\begin{tabular}{cll}
Level & [[5,1,3]]    & [[7,1,3]]   \\
\hline
0           & 6.30965616\% & 6.30965616\% \cr
1           & 6.29873094\% & 6.25921455\% \cr
2           & 6.29795843\% & 6.26714580\% \cr
3           & 6.29850925\% & 6.268(8)\%   \cr
4           & 6.299(0)\%   & 6.269(6)\%   \cr
5           & 6.299(3)\%   & 6.270(0)\%   \cr
6           & 6.299(5)\%   & 6.270(3)\%   \cr
7           & 6.299(6)\%   & 6.270(3)\%   \cr
$\infty$    & 6.299(6)\%   & 6.270(3)\%   \cr
\hline
Unoptimized & 4.58758548\% & 3.22981197\% \cr
\end{tabular}
\end{ruledtabular}
\end{table}

\begin{table}
\caption{\label{tab:separate} Critical values for a channel with independent probabilities $p$ of bit flip and phase flips $(p-p^2,p^2,p-p^2)$}.
\begin{ruledtabular}
\begin{tabular}{cll}
Level & [[5,1,3]] & [[7,1,3]] \\
\hline
0           & 11.00278644\% & 11.00278644\% \cr
1           & 10.94668310\% & 10.94286393\% \cr
2           & 10.94728109\% & 10.95683308\% \cr
3           & 10.949(1)\%   & 10.960(0)\%   \cr
4           & 10.949(9)\%   & 10.961(5)\%   \cr
5           & 10.950(4)\%   & 10.962(3)\%   \cr
6           & 10.950(7)\%   & 10.962(7)\%   \cr
7           & 10.950(8)\%   & 10.962(9)\%   \cr
$\infty$    & 10.951\%      & 10.963\%      \cr
\hline
Unoptimized & 7.14780025\%  &  6.45962393\% \cr
\end{tabular}
\end{ruledtabular}
\end{table}

\ignore{
\begin{table}
\caption{\label{tab:two} Critical values for $(p_X,p_Y,p_Z) = (p,0,p)$}
\begin{ruledtabular}
\begin{tabular}{cccc}
Level & [[5,1,3]] & [[7,1,3]] & 5 bit flip / 5 phase flip \\
\hline
0     & 11.35460976\%  & 11.35460976\% & 11.35460976 \% \cr
1     & 11.23967855\%  & 11.15332869\% & 11.33392680 \% \cr
2     & 11.23064987\%  & 11.16936898\% & 11.34181376 \% \cr
\end{tabular}
\end{ruledtabular}
\end{table}
}

\ignore{
\begin{table}
\caption{\label{tab:dom} Critical values for $(p_X,p_Y,p_Z) = (p,10^{-6},10^{-6})$. Note that for the $[[7,1,3]]$ code, the entropies are $\frac{1}{2}$.}
\begin{ruledtabular}
\begin{tabular}{cccc}
Level & [[5,1,3]] & [[7,1,3]] & 5 bit flip / 5 phase flip \\
\hline
0 & 49.62410483\% & 11.00138\% & 49.62410483\% \cr
1 & 49.64614794\% & 10.94281\% & 49.64614908\% \cr
2 & 49.66961046\% & 10.95678\% & 49.66961385\% \cr
\end{tabular}
\end{ruledtabular}
\end{table}
}

\section{Error correcting code thresholds}
In general, the thresholds of correctable noise for a code under optimized recovery operators can be estimated by analyzing for which physical noise the entropy of the resulting logical noise is equal to some specific value near the threshold.  Under repeated concatenation, these critical values converge to a threshold value. Generally, the entropy of the physical noise at the threshold is not $1$. Noise below or above the threshold tends toward a logical entropy of $0$ (no noise) or $2$ (totally depolarizing). At the threshold, the entropy at each level of the code varies a lot less than the average encoded error. This allows for more efficient calculations of the thresholds than using the error rate \cite{Poulin}. We disprove the conjecture \cite{Poulin} that depolarizing noise can be corrected up to an entropy of $1$ for the $[[5,1,3]]$ code.
 % allowing more efficient calculations of the thresholds than previous methods\cite{Poulin}. 
The critical values are shown in Tables \ref{tab:depolarizing} and \ref{tab:separate}. These results indicate that the first two levels, which were calculated exactly, provide a good estimate of the error thresholds. Higher levels were evaluated with Monte Carlo calculations; for more details see \cite{capacity}. In Tab. \ref{tab:depolarizing}, it can be seen tha,t for the depolarizing noise, the $[[5,1,3]]$ code has a threshold (level $\infty$) of $p=6.299(6)\%$, which is less than the physical noise with  an entropy of $1$ (level $0$) of $p=6.30965616\%$, disproving the conjecture of \cite{Poulin} that the threshold rate is equal to the latter.

The value of $p$ for which the channel Shannon entropy is $1$ was calculated for various levels of concatenation for two different types of noise: the depolarizing channel (Tab. \ref{tab:depolarizing}) and independent bit and phase flips (Tab. \ref{tab:separate}). Tables \ref{tab:depolarizing} and \ref{tab:separate} compare our thresholds to the unoptimized thresholds obtained from the channel maps in Refs. \cite{FKSS,Rahn:02a}.  The $p$ values converge to the channel threshold for correctable noise at the $\infty$ level.  The entropy at the thresholds for the $[[5,1,3]]$ code \cite{LMPZ} ranges from $0.993$ (for $(p_X,p_Y,p_Z)=(p,0,p)$ noise) to over $1$. For the $[[7,1,3]]$ code \cite{CSh}, both the phase flip noise $(0,0,p)$ and the channel $(p-p^2,p^2,p-p^2)$ have a critical value of $p=10.963\%$, yielding  minimum and maximum entropy of the thresholds as $0.4988$ and $0.9976$, respectively.

The results indicate that this adaptive concatenation approach greatly outperforms the concatenated channel map method. Tab. \ref{tab:depolarizing} shows that,  for depolarizing noise, the value of $p$ for the $[[7,1,3]]$ code is increased from $p=3.2298\%$ to $p=6.270(3)\%$. The $[[7,1,3]]$ code is particularly useful for quantum fault tolerance. It is therefore of interest to apply this adaptive concatenation method and similar entropy-based methods of threshold calculations to evaluate fault tolerance thresholds \cite{fault}.

For a discussion of which noise is correctable under the best code for correcting that noise, see \cite{capacity}.

\begin{acknowledgments}
We thank K. B. Whaley and M. Sarovar for reading and valuable criticisms of the manuscript. We thank the NSF for financial support under ITR Grant No. EIA-0205641.
\end{acknowledgments}

\ignore{

}
\ignore{

}
\end{document}